\documentstyle[prl,aps,epsf]{revtex}
\begin{document}

\newcommand{\si}{\sigma}

\twocolumn[
\hsize\textwidth\columnwidth\hsize\csname@twocolumnfalse\endcsname

\title{Comments on ``The Glassy Potts Model''}

\author{Domenico M. Carlucci}

\address{Department of Physics, 
Tokyo Institute of Technology, \\
Oh-okayama 
Meguro-ku
Tokyo 152-8551, 
Japan \\
{\rm e-mail: mimmo@stat.phys.titech.ac.jp}
}

\date{\today}
\maketitle
\begin{abstract}
We report the equivalence of the ``Glassy Potts model'', recently introduced 
by Marinari {\it et al.} and the ``Chiral 
Potts model'' investigated by Nishimori and Stephen. Both models  
do not exhibit any spontaneous magnetization at low temperature, 
differently from the ordinary glass Potts model. The phase  transition of 
the glassy Potts model is easily interpreted as the spin glass transition 
of the ordinary random Potts model. 
\end{abstract}
\twocolumn
\vskip.5pc ] 
\narrowtext


In a very recent paper Marinari {\it et al.}\cite{Marinari} introduced 
a $q$-component Potts model possessing the same gauge invariance of the Ising 
spin glass, thus inhibiting the presence of spontaneous magnetization  
at low temperature. Their model turns out to be a good candidate to describe 
the physical properties of the real glasses whose glassy phase must 
extend from $T_c$ down to $T=0$. In their Hamiltonian 

\begin{equation} 
   H\,=\, - \sum_{i,j} \delta_{\si_i; \Pi_{ij}(\si_j) }
   \label{Marinari_H}
\end{equation} 
the  spin variables take the values $\si_i=0,\,1,\,\dots,\,q-1$, and  
the role of the quenched disorder is played by the  random quenched 
{\it state permutations} $\Pi_{ij}$. Thus a non zero contribution to the 
Hamiltonian is given when $\si_i=\Pi_{ij}(\si_j)$, whereas in the 
standard  Potts model (see \cite{Binder} and references in \cite{Marinari})  
this is the case when the two spins are found in the same state 
$\si_i=\si_j$. 
Actually a gauge-invariant Hamiltonian (chiral Potts model) 
has been introduced many years ago by Nishimori and 
Stephen\cite{Nishimori}, although in a different form 
as generalization of Ising spin glass to the Potts glass 

   \begin{equation} 
     H\,=\,- \sum_{ij} \frac{1}{q} 
     \sum_{r=0}^{q-1} 
     J^{(r)}_{ij}
     \si_i^{r}\si_j^{q-r}
     \label{Nishimori_H} 
  \end{equation} 

where the spins are written in the complex representation (each spin 
has two components and $\left(\si_i^r\right)^*=\si_i^{q-r}$), 
each state being represented by one of  the $q$ 
roots of unity,  and $J_{ij}$ are complex random quenched 
variables. The additional condition on the coupling constants 
$\left(J_{ij}^{(r)}\right)^*=J_{ij}^{(q-r)}$
insures the realness of the expression (\ref{Nishimori_H}). Of course, 
when $q=2$, the Hamiltonian (\ref{Nishimori_H}) reduces to the well know 
Ising spin glass problem. Moreover it is worthy to be noticed that 
when the $\{J\}$'s are constant, one recovers 
the usual non random Potts model ($H=-J \sum_{ij}\delta_{\si_i;\si_j}$), 
because of the following formula 

\begin{equation} 
   \delta_{\si_i;\si_j} 
   \,=\, 
   \frac{1}{q} 
   \sum_{r=0}^{q-1}   
   \si_i^{r}\si_j^{q-r}
   \label{formula}
\end{equation}

If one considers a discrete distribution of the coupling constant 
 \begin{equation} 
   J^{(r)}_{ij} =  \tau^r_{ij}, 
\end{equation} 
$\{\tau\}$ being a root of the unity associated to the link 
$\langle ij\rangle$with some probability weight, 
it is straightforward to check that one recovers the Hamiltonian 
(\ref{Marinari_H}), since the Hamiltonian differs from zero when 
$\si_i=\tau_{ij} \si_j$, $\tau_{ij}$ acting as a random permutation of 
the spin values. In the mean field limit, the chiral Potts 
model can be easily solved by means of the replica trick\cite{Mezard}, 
thus leading at the minimization of the following free energy density

\[ 
  \hspace{-1cm}
  -\beta f
  \,=\, 
  - 
  \frac{J_0 \beta }{2 q }\sum_{\alpha}\sum_{r=1}^{q-1} 
  \left[ (m_{1;r}^{\alpha})^2 + (m_{2;r}^{\alpha})^2\right] 
   -
\]
\[
  \,-\,
   \frac{1}{2} \left(\frac{J \beta}{q}\right)^2
   \sum_{(\alpha\beta)}\sum_{r=1}^{q-1} 
    \left[ (Q_{1;r}^{(\alpha\beta)})^2 + (Q_{2;r}^{(\alpha\beta)})^2 \right]
   \,+\,
\]
\[
 \,+\, 
\log {\rm Tr}\exp\left\{ 
    \left(\frac{J \beta}{q}\right)^2
    \sum_{(\alpha\beta)}
    \sum_{r=1}^{q-1} 
     \left[ Q_{1;r}^{(\alpha\beta)} 
			{\rm Re} \left[  
                                   (\si^{\alpha})^{r} 
                                   (\si^{\beta})^{q-r}
                             \right] 
                     \,+\, \right.\right.
\]
\[ 
 \left.\left.
  \,+\,
             Q_{2;r}^{(\alpha\beta)} 
			{\rm Im} \left[  
                                   (\si^{\alpha})^{r} 
                                   (\si^{\beta})^{q-r}
                             \right] 
      \right]
       \,+\,
                  \right. 
\]
\begin{equation} 
\left.
  \,+\,
   \left(\frac{J_0 \beta}{q}\right)
    \sum_{\alpha}
    \sum_{r=1}^{q-1} 
     \left[ m_{1;r}^{\alpha} 
			{\rm Re} \left[  
                                   (\si^{\alpha})^{r} 
                             \right] 
                     \,+\, 
             m_{2;r}^{\alpha} 
			{\rm Im} \left[  
                                   (\si^{\alpha})^{r} 
                                  \right]
      \right]
\right.
\label{free_energy}
\end{equation}

It is straightforward to check that for $q=2$, the imaginary 
parts in (\ref{free_energy}) vanishes and one obtains the 
replicated SK free energy. Here the coefficient of the magnetization 
is simply $J_0 \beta/q$ to be compared with the coefficient one gets 
in the standard Potts model\cite{Binder} $\beta\left( J_0 + \frac{1}{2}(q-2) \beta J^2\right)$, responsible for the ferromagnetic phase at low temperature 
in absence of magnetic field.
In order to investigate the para-glass (isotropic) transition, where 
the magnetization vanishes $m_{1;r}^{\alpha}=m_{2;r}^{\alpha}=0$, we assume 
that the spin glass order parameters do not depend on $r$. This implies 
$Q_{1;r}^{(\alpha\beta)}= Q^{(\alpha\beta)}$ and 
$Q_{2;r}^{(\alpha\beta)}= 0$. Therefore, from (\ref{formula}) and 
(\ref{free_energy}) the free-energy to be minimized simply reads 

\[ 
-\beta f\,=\, -\frac{1}{2} (q-1) \left( \frac{J\beta}{q}\right)^2
\sum_{(\alpha\beta)} \left(Q^{(\alpha\beta)}\right)^2 \,+\, 
\]
\begin{equation} 
\,+\, 
\log{\rm Tr}\exp\left[ 
                       \left( \frac{J\beta}{q}\right)^2
                       \sum_{(\alpha\beta)} Q^{(\alpha\beta)} 
\left( \delta_{\si^{\alpha} \si^{\beta}} -\frac{1}{q} \right) 
\right]
\end{equation}
              
which turns out to be exactly the same free energy of the standard random 
Potts model\cite{Binder}, with the practical advantage 
that the glass phase extends down to $T=0$. Therefore for $q=4$, 
the transition from the paramagnetic phase 
to the glass phase is discontinuous with one step replica symmetry breaking
at least above the upper critical dimension. 
On the other hand, the numerical calculations performed in \cite{Marinari} seem to indicate 
that the REM-like landscape of the free energy holds also in d=4. 

\vspace{1cm}

This work is supported by the JSPS under Grant No. P96215.
                
The author warmly thanks H.Nishimori for stimulating discussions 
and E.Marinari and G. Parisi for useful correspondence.




\begin{references}

\bibitem{Marinari} 
 E.~Marinari, S.~Mossa and G.~Parisi
  cond-mat/9805300


\bibitem{Nishimori} 
H.~Nishimori and M.~J.~Stephen, 
Phys. Rev. B {\bf 27}, 5644 (1983)


\bibitem{Mezard} 
  M. Mezard, G. Parisi and M.A. Virasoro, 
  {\em Spin Glass Theory and Beyond}
  (World Scientific, Singapore, 1987).


\bibitem{Binder} 
  K. Binder,
  in  {\em Spin Glasses and Random Fields},
  edited by A.~P.~Young (World Scientific, Singapore 1997). 


\end{references}
\end{document}